\documentstyle[12pt,amssymb,a4,psfig]{article}
\begin{document}

\begin{center}
{\Large {\bf
Orbital Angular Momentum Parton Distributions in Light-Front Dynamics
}}
\end{center}

\vspace{1cm}

\begin{center}
{\large F. Cano$^1$, P. Faccioli$^{2,3}$, S. Scopetta$^{2}$ and  M. Traini$^{1,4}$}
\end{center}

\begin{center}
$^1$Dipartimento di Fisica \\
Universit\`a degli Studi di Trento \\
I-38050 Povo, Italy.

\vspace{0.2cm}
$^2$ ECT*,\\
European Centre for Theoretical Studies \\ 
in Nuclear Physics and Related Areas,\\
Villa Tambosi, I-38050 Villazzano (Trento), Italy.

\vspace{0.2cm}
$^3$ Department of Physics and Astronomy, \\ State
University of New York at Stony Brook, USA. 

\vspace{0.2cm}
$^4$ Istituto Nazionale di Fisica Nucleare, G.C. Trento.
\end{center}

\vspace{2cm}

\begin{abstract}
{\small 
We study the quark angular momentum distribution in the nucleon within
a light-front covariant quark model. Special emphasis is put into the
orbital angular momentum: a quantity which is very sensitive to the
relativistic treatment of the spin in a light-front dynamical
approach.  Discrepancies with the predictions of the low-energy
traditional quark models where relativistic spin effects are
neglected, are visible also after perturbative evolution to higher
momentum scales.  Orbital angular momentum distributions and their
contribution to the spin sum rule are calculated for different
phenomenological mass operators and compared with the results of the
MIT bag model.}
\end{abstract}

\vspace{0.2cm}

%\vspace{1cm}
%\noindent 
%cano@science.unitn.it \\
%traini@science.unitn.it \\
%faccioli@science.unitn.it

\vfill
%\hfill
\noindent
{\it Preprint} {\bf UTF-438,ECT*-00-002}
%\vfill
%\hfill
%{\it Preprint} {\bf }
 
\newpage

\section{Introduction}

	The EMC measurement of the integrated helicity parton
distributions \cite{EMC88} (for recent results at SLAC see
\cite{SMC94}) triggered the interest in a deeper understanding of how
the total angular momentum of the nucleon is shared among its
constituents.  It was found that the fraction of the nucleon spin
carried by the quarks was rather small, at variance with the most
naive quark model expectation where the proton spin is (almost)
entirely built from the spin of the quarks. Among the different
explanations of these discrepancies, it was proposed
\cite{ALTARELLI88} that the (overlooked until then) polarization of
the gluons might also contribute to the singlet axial charge through
the axial anomaly. In that case, experimental data would be compatible
with a rather large fraction of the spin carried by the quarks
($\Delta \Sigma = 0.45 \pm 0.09$ in a recent world data analysis
\cite{ALTARELLI97}), though still far away from the non-relativistic
quark model predictions.

 One of the most important issues raised by the 'spin crisis' was the
need for considering all the possible sources of angular momentum in
the nucleon. Therefore, the spin sum rule should read \cite{SEHGAL74,JAFFE90}:

\begin{equation}
\frac{1}{2} \Delta \Sigma(Q^2) + \Delta g (Q^2) + L_q(Q^2) +
L_g({Q^2})=\frac{1}{2}\;,
\label{spinsumrule}
\end{equation}	

\noindent where $\frac{1}{2} \Delta \Sigma(Q^2)$ ($\Delta g (Q^2)$) is
the spin carried by the quarks and antiquarks (gluons) and $L_q(Q^2)$
($L_g(Q^2)$) is the orbital angular momentum (OAM) contribution of the
quarks (gluons) \cite{LAMPE98}.

The significant role of OAM was pointed out several years ago
\cite{SEHGAL74,RATCLIFFE87}, but the problem was rigorously formulated only
recently, when a gauge invariant definition of the quark and gluon
(twist-two) operators was proposed \cite{JI97,ja99}.  Besides, there
has been a big effort to derive evolution equations (at one-loop
level) for OAM observables \cite{JI97,JI96}. At the present there is
only one gauge invariant definition of quark OAM \cite{JI97} with
known $Q^2$ evolution and experimentally accessible (for a discussion
cfr. Ref. \cite{ji99}). Such a definition holds for reference frames
with definite nucleon polarization and the OAM distribution could be
measured through the forward limit of skewed parton
distributions.
 
One peculiar feature, already expected by general arguments
\cite{SEHGAL74,RATCLIFFE87}, is explicitly realized in the evolution
equations, namely that
the OAM distribution is coupled to the helicity parton distributions.
As a consequence, OAM contributions can be
generated, through evolution at higher scales, even in the case of
vanishing OAM components at low initial hadronic scale.  This is
indeed the case for most hadronic models where the quarks are
arranged, in the ground state, in a $l=0$ S-wave configuration,
stressing the crucial role of $Q^2$ evolution for the evaluation of
spin observables or, in other words, the roughness of the
identification of constituent quarks with partons at all energy
scales.

	From a quantitative point of view, some studies are currently
	available \cite{SCOPETTA99,MA98,MA93}. In particular, in
	ref. \cite{SCOPETTA99}, the OAM distributions have been
	calculated for a number of hadronic models.  As a first step
	these quantities are evaluated at the hadronic (low energy)
	scale and then evolved to the experimental $ Q^2 >> \mu_0^2$
	scale by using the Leading Order (LO) evolution procedure
	recently established \cite{JI97}. 
	One central conclusion of that work is
	that a sizeable initial OAM distributions can deeply influence
	the final high-energy results.  As a consequence, a clear
	difference arises between non-relativistic and relativistic
	models: while the former usually give a tiny OAM contribution
	at $\mu_0^2$, the latter may give rise to sizeable effects at
	high $x$ that persist after evolution. To this respect, OAM
	distributions are useful quantities to assess the relevance of
	relativistic effects in the hadronic models of the nucleon.

In a recent study we investigated the consequences of a light-front
treatment of relativistic spin effects on the helicity distributions
\cite{FACCIOLI99,CANO99} and in the present paper we want to enlarge
our analysis to OAM investigating in detail the predictions of a
light-front covariant quark model. As a matter of fact, the spin
dynamics can be discussed within the light-front approach in a way
which respects covariance requirements and particularly suitable to
discuss deep inelastic polarized processes, both at the hadronic
\cite{MA93,FACCIOLI99,CANO99} and high-energy (partonic) scale
\cite{FACCIOLI99,CANO99}.  We will show that light-front covariant
quark models (LFCQM) predict a non-vanishing OAM distribution whose
main features survive after evolution.  We will also see that these
predictions hold for a variety of mass operators indicating that the
relevant ingredient is the relativistic treatment of the spin wave
functions, absent in many traditional formulations of the quark
model. The comparison with other relativistic models (MIT bag model) and
the analysis of the moments that enter the spin sum rule will allow
us to assess the reliability of LFCQM.

\section{OAM at the hadronic scale}

	In the recent years a quark model-based approach has been
developed for computing the non-perturbative inputs in the evolution
equations \cite{METHOD1} and describing polarized and unpolarized
parton distributions. Schematically, the central assumption is that at
some low-energy scale ($\mu_0^2$) the nucleon is made up of valence
quarks that can be identified with the constituents of the quark model
(or the bag model in alternative treatments). Therefore, the
non-perturbative boundary conditions can be evaluated by using
low-energy models of the nucleon. Subsequent refinements led to the
inclusion of non-perturbative gluons and sea at the hadronic scale
$\mu_0^2$ \cite{METHOD2}, as well as the explicit partonic content of
the constituent quarks \cite{METHOD3}.

	By following such a procedure we will assume that at the
hadronic scale only valence quarks are resolved so that the quark
helicity distribution $g_1^a(x,\mu_0^2)$ for a given flavour $a$ is
given in terms of the momentum density of the valence quarks:

\begin{equation}
g_1^a(x,\mu_0^2) = \frac{1}{(1-x)^2} \int \; d\vec{k} \;
(n_a^{\uparrow}(\vec{k}\,) - n_a^{\downarrow}(\vec{k}\,)) \; \delta
\left( \frac{x}{1-x} - \frac{k^+}{M_N} \right)\;,
\label{g1a}
\end{equation}	
  
\noindent 
where $x$ is the Bjorken variable, $M_N$ the mass of the nucleon and
$k^+$ is defined as a function of the parton momentum as
$k^+=\sqrt{\vec{k}^{\: 2} + m^2} + k_z$. The polarized momentum
densities are defined as:

\begin{equation}
n_a^{\uparrow \; \downarrow}(\vec{k}) = \langle N, J_z=1/2 |
\sum_{i=1}^3 {\cal{P}}_a \frac{1 \pm \sigma^{(i)}_z}{2} \delta(\vec{k}_i -
\vec{k}) | N, J_z=1/2 \rangle\;,
\end{equation}

\noindent where ${\cal{P}}_a$ is the flavour projector. 

An analogous definition can be worked out for the OAM distributions
\cite{SCOPETTA99}:

\begin{equation}
L_z(x,\mu_0^2) = \frac{1}{(1-x)^2} \int \; d\vec{k} \; L_z(\vec{k}\,)
\; \delta \left( \frac{x}{1-x} - \frac{k^+}{M_N} \right) \;,
\label{Lzx}
\end{equation}

\noindent
where the density of the angular momentum is defined in the usual way:

\begin{equation}
L_z(\vec{k}) = \langle N, J_z=1/2 | \sum_{i=1}^3 - i (\vec{k}_i \times
\vec{\bigtriangledown}_{\vec{k}_i}\:)_z \delta(\vec{k}_i - \vec{k}) |
N, J_z=1/2 \rangle \;.
\label{lzk}
\end{equation}

	From the previous definitions one can recover the result
$L_z(x,\mu_0^2)=0$ obtained assuming S-wave quarks only, and
non-relativistic approximation (i.e. $L_z(\vec{k}\,)=0$).  A more
complicated example within non-relativistic dynamics, is given
by models whose nucleon wave function is
a superposition of various $SU(6)$ components, such as  
the Isgur-Karl model \cite{ISGUR78} . The non-vanishing
contribution to $L_z(\vec{k}\,)$ in these cases is due to the D-state
(or higher waves) admixture. For example, in the Isgur-Karl model, the
OAM distribution of Eq. (\ref{Lzx}) results to be proportional to the
D-State probability $a_D^2$ \cite{SCOPETTA99} and its contribution is
very small ($a_D=-0.067$).

	This situation is radically changed in a light-front covariant
quark model.  In light-front dynamics (LFD) \cite{KEISTER91}, the
specific partition of the Poincar\'e algebra into kinetic and
Hamiltonian generators leads to several simplifications of the
relativistic many-body problem such as, for example, the clean
separation of the center of mass motion.  The prize to pay is that the
description of angular momentum is rather complicated.  Not all the
generators of rotations, in fact, belong to the kinetic subgroup, and
hence the angular momentum operator is, in general, interaction
dependent.\\ For this reason, in the phenomenological applications of
LFD to the quark model, it is customary to work in the
Bakamjian-Thomas construction \cite{KEISTER91}, that is adding a
phenomenological interaction to the free mass operator, only.
However, the resulting total angular momentum operator, although
interaction free, does not satisfy ordinary composition rules.  In
order to restore them, a unitary transformation of the Hilbert space,
known as Melosh Rotation (MR), has to be performed.  In particular, if
the nucleon is in a S-wave state, such rotation acts only on the spin
part of the wave function.

The $D^{1/2}$ representation of the MR is given by:

\begin{equation}
D^{1/2}[R_M(\vec{k})] = \frac{(m+\omega+k_z) - i
\vec{\sigma}\cdot(\hat{z} \times \vec{k}_\perp)}{( (m+\omega+k_z)^2 +
\vec{k}_\perp^{\,2})^{1/2}} \;\; .
\label{mr}
\end{equation}    

	As a result, motion and spin are now intimately correlated as
it is required by a relativistic theory. The MR can be interpreted as
the boost transformations required to move from the rest frame of each
subsystem (quark) to the rest frame of the total system (nucleon).

In the present study we will not investigate $SU(6)$ breaking effects
in spin-isospin space: the nucleon wave function will correspond to a
S-wave.  This simplifying assumption ensures that the non-vanishing OAM
contribution originates from pure relativistic effects due to the
treatment of the spin in light-front dynamics.  Indeed the MR gives
rise to a non-vanishing angular momentum density even if the spatial
wave function corresponds to a S-wave and the angular momentum density
can be written, for a $SU(6)$ symmetric spin-isospin wave function, as

\begin{equation}
L_z(\vec{k}\,) = \frac{1}{3} \frac{\vec{k}_\bot^{\,2}}{ (m + \omega +
k_z)^2 + \vec{k}_\bot^{\,2}}\, n(\vec{k}\,)\;,
\label{LzLF}
\end{equation} 

\noindent 
where $n(\vec{k}\,)$ is the total momentum density, defined in the
usual way :

\begin{equation}
n(\vec{k}\,) = \langle \Psi_N | \sum_{i=1}^3 \delta(\vec{k}_i -
\vec{k}\,) | \Psi_N\rangle \;,
\end{equation}

\noindent and normalized to the number of particles ($\int n(\vec
k\,)\,d\vec k = 3$). Recalling the expressions for the polarized
densities that enter the helicity distributions
\cite{FACCIOLI99,CANO99}:

\begin{eqnarray}
n_u^{\uparrow}(\vec{k}\,) - n_u^{\downarrow}(\vec{k}\,) & = & - 4\,
\left(n_d^{\uparrow}(\vec{k}\,) - n_d^{\downarrow}(\vec{k}\,)\right)
\nonumber \\ & = & \frac{4}{9} \frac{(m+\omega+k_z)^2 -
\vec{k}_\perp^{\,2}}{(m+\omega+k_z)^2 + \vec{k}_\perp^{\,2}}\,n(\vec
k\,) \;,
\label{DnLF}
\end{eqnarray}

\noindent one can check that the total angular momentum sum rule is
automatically fulfilled at the hadronic scale:

\begin{equation}
\frac{1}{2} \int (g_1^u(x,\mu_0^2) + g_1^d(x,\mu_0^2)) \; dx + \int \;
dx L_z(x,\mu_0^2) = \frac{1}{2}\;.
\end{equation}

Another interesting relationship connects $L_z$ to the longitudinal
($g_1$) and transversity ($h_1$) parton distributions, namely
\cite{MA98}:

\begin{equation}
g_1^a(x,\mu_0^2) + L_z^a(x,\mu_0^2) = h_1^a(x,\mu_0^2)\;,
\label{MArel}
\end{equation}

\noindent
and is naturally fulfilled in our approach. This relationship also
holds for other relativistic models of the nucleon, such as the bag
model. Let us stress that Eq. (\ref{MArel}) is valid at the hadronic
scale $\mu_0^2$ only and one should be careful when using it to
extract information about $L_z^a$ because it is broken by evolution,
even at small values of $Q^2$. This can be easily demonstrated by
considering the singlet combination corresponding to
Eq. (\ref{MArel}), i.e.

\begin{equation}
\Sigma(x,\mu_0^2) + L_z(x,\mu_0^2) - H(x,\mu_0^2) = 0
\label{MAsinglet}
\end{equation} 

\noindent
where $\Sigma(x,\mu_0^2) = \Sigma_a (g_1^a(x,\mu_0^2) +
g_1^{\bar{a}}(x,\mu_0^2))$ and $H(x,\mu_0^2) = \Sigma_a
(h_1^a(x,\mu_0^2) - h_1^{\bar{a}}(x,\mu_0^2) )$\footnote{The minus
sign in front of $h_1^{\bar{a}}$ comes from the properties of the
operator that defines the transversity under charge conjugation, and
therefore the analogous of Eq. (\ref{MArel}) for antiquarks should
read $g_1^{\bar{a}}(x,\mu_0^2) + L_z^{\bar{a}}(x,\mu_0^2) = -
h_1^{\bar{a}}(x,\mu_0^2)$. Though our model does not contain
antiquarks at the scale $\mu_0^2$, this relationship can be easily
checked in the bag model.}. In order to check the validity of
Eq. (\ref{MArel}) at $Q^2 > \mu_0^2$ let us evolve (at LO) the first
moments of the left-hand side of Eq.  (\ref{MAsinglet}):

\begin{eqnarray}
\langle \Sigma(x,Q^2)\rangle_1 + \langle L_z(x,Q^2)\rangle_1 - \langle
H(x,Q^2) \rangle_1 & = & \frac{1}{2} (1-b^{-50/81}) \langle
\Sigma(x,\mu_0^2)\rangle_1 \nonumber \\ & + & (b^{-50/81}- b^{-4/27})
\langle H(x,\mu_0^2) \rangle_1 \nonumber \\ & + & \frac{9}{50}
(1-b^{-50/81})
\end{eqnarray}

\noindent
where $b=\frac{\ln(Q^2/\Lambda^2)}{\ln(\mu_0^2/\Lambda^2)}$. Clearly,
the right-hand side of the equation above vanishes only if
$Q^2=\mu_0^2$. Furthermore, due to the form of the $b$-dependent
coefficients, it quickly deviates from 0 at the initial stages of
evolution pointing out the limits of the attempt, carried out in
\cite{MA98}, of extracting information on $L_z$ from 
Eq. (\ref{MArel}).

	Coming back to our evaluation of OAM, Eqs. (\ref{LzLF}) -
(\ref{DnLF}) show that the exact ratio between the amount of OAM and
spin will depend on the specific form of $n(\vec{k}\,)$, or
equivalently, on the spatial nucleon wave function. Let us note
however that the momentum density averages many details of the spatial
wave function and to this respect, the sensitivity of the final
results to the fine details of the spatial wave function is reduced.
 
In the following we will discuss predictions obtained solving
explicitly the mass equation:

\begin{equation}
\hat M \, \Psi = (\sum_{i_1}^3 \sqrt{\vec{k}_i^{\, 2} + m^2} + V)\,
\Psi = E \,\Psi
\label{mass1}
\end{equation}

\noindent 
with an hypercentral phenomenological potential:

\begin{equation}
V= - \frac{\tau}{\xi} + \kappa_l \xi + \Delta \;\; ,
\label{mass2}
\end{equation} 

\noindent 
where $\xi$ is the hyperradius defined in the usual way and $\tau$,
$\kappa_l$ and $\Delta$ are free parameters that are fixed by
spectroscopy requirements \cite{FACCIOLI99,PAOLO99,TBM95}. It is worthwhile
mentioning that MR has no effects on the energy levels of the
confining mass operator (\ref{mass1}) - (\ref{mass2}) explaining to
some extent the success of non-relativistic (or relativized)
approaches in reproducing the baryonic spectrum. On the other hand,
another remarkable effect of the relativistic mass equation is the
enhancement of the high momentum components in the nucleon wave
function. Since the MR factor involves momentum dependent terms, the
final results will be biased by the presence of these high-momentum
components. In order to test the sensitivity to the details of the
momentum density we will consider an additional scenario where the MR
factors are combined with a wave function obtained from the
non-relativistic Schr\"odinger reduction of the Eq. (\ref{mass1}) with
the same form of potential (\ref{mass2}). 
This new spatial wave function, hereafter indicated by $\Psi'$,
will contain far less high momentum components. In fact one of the
risks of guessing the wave function instead of solving the mass
equation (\ref{mass1}) explicitly, is to underestimate the contribution
coming from the high-momentum components of the correct solution,
mostly carried over by the relativistic kinetic energy operator in the
mass equation.  Although the use of MR is not fully consistent when
$\Psi'$ is considered, since it was derived from a non-relativistic
mass equation, we will discuss it as a 'pedagogical' example that
represents an extreme scenario where high-momentum components have
been strongly suppressed. The comparison of results obtained with
$\Psi$ and $\Psi'$ will serve to establish bounds on the effects of
MR.

\section{Results and discussion}

	The obtained OAM distribution at the hadronic scale $\mu_0^2$,
Eq. (\ref{Lzx}), 
for the wave function $\Psi$, the solution of Eq. (\ref{mass1}), 
is shown in Fig. 1.a. The outcome for the
modified scenario (corresponding to the wave function $\Psi'$) is
shown in Fig. 1.b to appreciate the effect of the lack of high
momentum components. Furthermore, the comparison with the bag model
results \cite{SCOPETTA99} is also provided (Fig. 1.c). It is clear that the LFCQM,
regardless of the details of the spatial wave function, provides OAM
distributions which are comparable (even bigger by a factor 2) to the bag
model. From the comparison between Figs. 1.a and 1.b one can see that
the MR (and not the specific shape of the spatial wave function) is
responsible for this sizeable OAM. In non-covariant quark models such
as the Isgur-Karl model, where MR is omitted, the OAM distributions is
almost flat \cite{SCOPETTA99}.  Even when considering a D-model
\cite{ROPELE95} where the probability of the D-wave component is
raised up to a 20 \%, the resulting OAM, though comparable in size to
those obtained here, are peaked at lower $x$. Nonetheless the large
deformation of the nucleon in the D-model should not be taken as
realistic.

	In order to bring the OAM distributions to the high-energy
experimental scale, we use the recently obtained evolution equations
at LO \cite{JI97,MARTIN99}. In the process the OAM distributions for the gluons
will be generated. The initial scale $\mu_0^2$ is determined following
the criteria exposed in \cite{FACCIOLI99}, and at LO turns out to be
$\mu_0^2=0.079$ GeV$^2$. In fig. 1.a and 1.b we also present the
evolved OAM distributions up to $Q^2=10$ GeV$^2$ (short-dashed line)
and $Q^2=1000$ GeV$^2$ (long-dashed line).

By comparing again the LFCQM with the bag model (Fig. 1.c) it is clear
that a non-vanishing OAM persists in the large $x$ region and this is
a distinctive feature of relativistic treatments of the nucleon.
Indeed, in I-K models, the OAM is entirely concentrated at low $x$.
This may constitute a clear signature of relativity in the low-energy
models of the nucleon if $L_z(x,Q^2)$ is measured.

	In our approach all the gluon OAM is generated through
	evolution. In Fig. 2 we present the resulting $L_g(x,Q^2=10 \;
	\mbox{GeV}^2)$ for $\Psi$, $\Psi '$ and the bag model. There
	is an inverse correlation between the amount of high momentum
	components in the wave function and the value of $L_g$ at
	small $x$. The OAM gluon distribution for the bag model falls
	between those obtained with $\Psi$ and $\Psi '$.

	Concerning the first moments of the distributions, our model
gives a value for $L_q(\mu_0^2) = \int L_z(x,\mu_0^2) \; dx$ that ranges
from 0.272 to 0.126 ($\Psi$ and $\Psi'$ model respectively). It should
be stressed that the corresponding values for $\Delta \Sigma$ (0.456
and 0.748 respectively) are {\it per se} a clearcut signature in favor
of light-front quark models, when compared to recent analysis of data
($\Delta \Sigma = 0.45 \pm 0.09$) \cite{ALTARELLI97}.
Furthermore, these numbers are quite close to the angular momentum
share-out given by the bag model.

	The first moments that make up the spin sum rule also evolve
with $Q^2$ according to \cite{MARTIN99}:

\begin{eqnarray}
\frac{1}{2} \Delta \Sigma (Q^2) & = & \frac{1}{2} \Delta
\Sigma(\mu_0^2) \\ L_q (Q^2) & = & (b^{-\frac{50}{81}} -1) \frac{1}{2}
\Delta \Sigma(\mu_0^2) + b^{-\frac{50}{81}} L_q(\mu_0^2) -\frac{9}{50}
(b^{- \frac{50}{81}}-1) \\ J_g (Q^2) & = & b^{-\frac{50}{81}}
J_g(\mu_0^2) -\frac{8}{25}(b^{- \frac{50}{81}}-1)
\end{eqnarray}

	In Fig. 3 we show the evolution of these quantities with $Q^2$
for $\Psi$ (Fig. 3.a) and $\Psi '$ (Fig. 3.b).  It is worthwhile
mentioning that, even if we do not have gluons at the hadronic scale,
they quite rapidly develop a sizeable angular momentum content.
Furthermore the gluon angular momentum evolves decoupled from the
quark sector and if we start with a vanishing $J_g(\mu_0^2)$ then
$J_g(Q^2)$ is completely determined by the QCD anomalous
dimensions. The values for $J_g(Q^2)$ in the region between 1 and 10
GeV$^2$ that we find ($J_g \sim 0.20-0.25$) are compatible with those
found by using QCD sum rules \cite{BALITSKY97} ($J_g \sim 0.25$) and in
a recent lattice calculation \cite{MATHUR99} ($J_g=0.20 \pm
0.07$). They also agree with another model calculation based on the
one-gluon exchange interaction between quarks \cite{BARONE98} ($J_g
\sim 0.24$). The consideration of a non-vanishing $J_g(\mu_0^2)$ would
not change much our results since it would also raise the scale
$\mu_0^2$ (due to the fact that at that scale the gluons would carry
some momentum) and hence $b$ would be larger for a given $Q^2$.

Though the large error bars in the first direct measurement of the
ratio $\Delta g \over g$ \cite{HERMES99},  
${\Delta g \over g} = 0.41 \pm 0.18 (stat) \pm 0.03 (syst)$, and the
values for $ \Delta g$ obtained in recent data analysis
\cite{ALTARELLI97}, $\Delta g(Q^2=1 \mbox{ GeV}^2) = 1.6 \pm 0.9$, 
do not allow to discriminate between
models, our results fall within the range of the latter. 
As a matter of fact the
rather moderate values for $J_g$ result from a strong cancellation
between $L_q(Q^2)$ and $\Delta g(Q^2)$ and, in particular we have
$\Delta g(Q^2=1 \mbox{ GeV}^2) = $ 1.36 and 2.22 for $\Psi$ and
$\Psi'$ respectively.

	In a non-relativistic quark model one would expect $L_q(Q^2)
	\sim - J_g(Q^2) \sim - 0.25$ in the range $Q^2 \sim 1 - 10$
	GeV$^2$ since $\Delta \Sigma$ is a constant with $Q^2$. When
	relativistic spin effects are taken into account, as Fig. 3
	shows, one expects $L_q(Q^2)$ to be much smaller ($L_q(Q^2)
	\sim - 0.12$ at most) or close to zero.

\section{Concluding remarks}

	In summary we have shown that covariant light-front based
quark models give rise to non-trivial predictions for the OAM
distributions at both low and high momentum scales. This departure
from traditional treatments of the angular momentum structure of the
nucleon is more manifest in the high-$x$ region of the quark sector. 
We have seen
that the performance of LFCQM is quite similar to other relativistic
models of the nucleon such as the bag model. This comparison holds for
a quite flexible choice of the mass operator. We have studied the
predictions for other potentials that interpolate between the two
somehow extreme situations presented here and conclusions are not
changed. In fact, there is a clear correlation between the amount of
high-momentum components in the momentum density $n(\vec{k})$ and the
size of the OAM distribution. A more realistic interaction would give
results closer to those of $\Psi$ than to those obtained with $\Psi '$
because a relativistic treatment of the kinetic energy operator
inevitably emphasizes the high-momentum tail.

	One should keep in mind however that the origin of the
relativistic aspects is not the same in the bag and in the LFCQM
presented here. While in the former the non-vanishing OAM comes from
the small Dirac components, in the latter these ones are absent and
relativity enters through the momentum dependence of the Pauli
spinors. Certainly other more sophisticated spin-flavour basis can be
constructed, such as the Dirac-Melosh bases \cite{BEYER98}, where
covariance is manifest. Despite the fact that the used basis contains
only kinematic and not dynamical (higher Fock states) effects, it
represents a minimal framework that combines in an elegant way
simplicity and a proper treatment of boost.  Results obtained with
this basis and with the bag model are of similar quality pointing out
that it allows an easy implementation of relativistic effects in the
spin structure of the nucleon.

\vskip 0.5cm

We gratefully acknowledge Vicente Vento for useful comments 
and a careful reading of the manuscript.

\newpage 

\begin{center}
{\bf Figure captions}
\end{center}

\begin{description}

\item{\bf Figure 1} Quark orbital angular momentum distributions
calculated in light-front dynamics with the wave function $\Psi$ (a),
with the modified wave function $\Psi '$ (see text) (b) and in the bag
model (c). Solid lines correspond to the initial hadronic scale
$\mu_0^2$, short-dashed lines to $Q^2=10$ GeV$^2$ and long-dashed ones
to $Q^2=1000$ GeV$^2$.

\item{\bf Figure 2.} Gluon orbital angular momentum distributions
calculated at $Q^2=10$ GeV$^2$ with the wave function $\Psi$ (solid
line), $\Psi'$ (long-dashed line) and the bag model (short-dashed
line).

\item{\bf Figure 3.} The contributions to the proton spin sum rule
according to the model with $\Psi$ (a) and $\Psi'$ (b). The dashed
curve shows $\frac{1}{2} \Delta \Sigma(Q^2)$, the long-dashed one is
$L_q(Q^2)$ and the dot-dashed curve represents $J_g(Q^2)$.

\end{description}

\newpage
 
\
\vfill
\centerline{\protect\hbox{\psfig{file=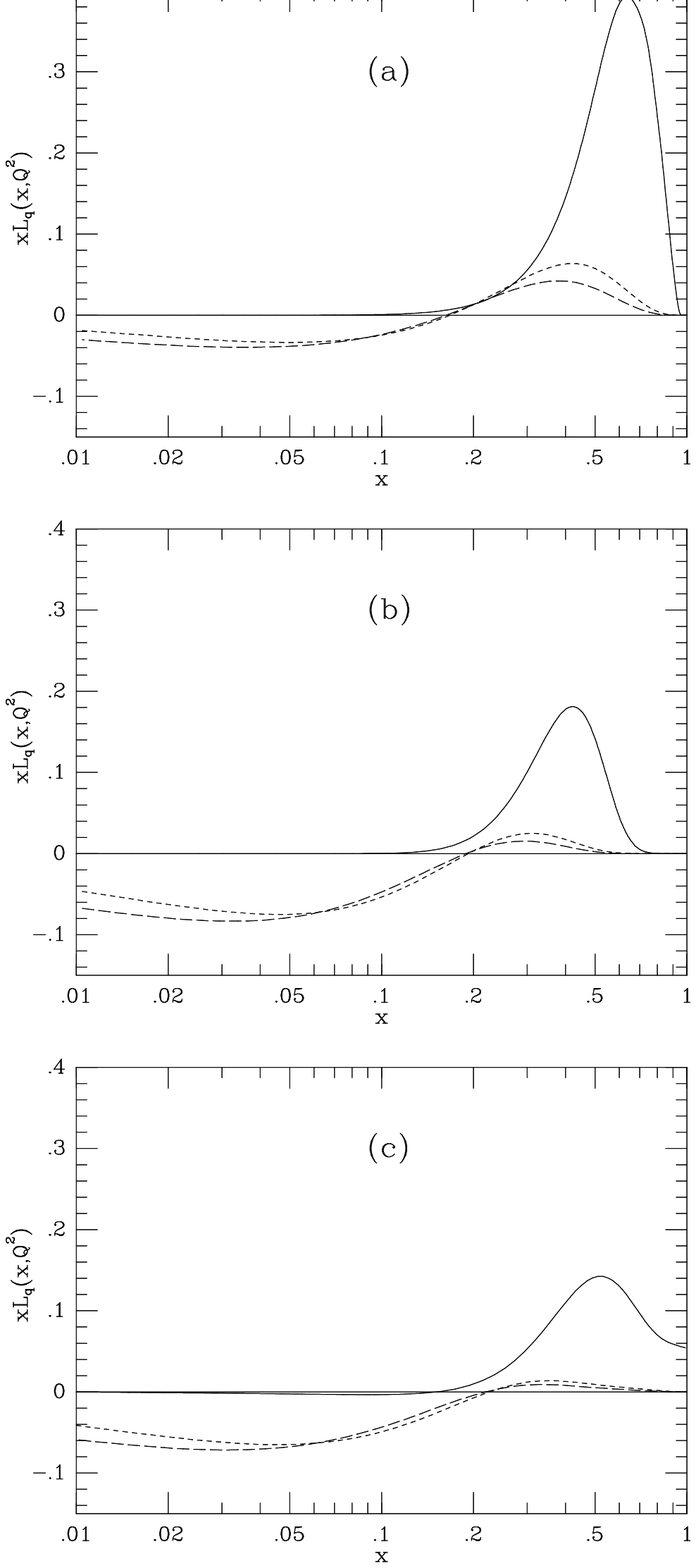,width=0.98\textwidth}}}
\vskip 1cm
\centerline{\bf Figure 1}
\vfill
%\newpage
\
\vfill
\centerline{\protect\hbox{\psfig{file=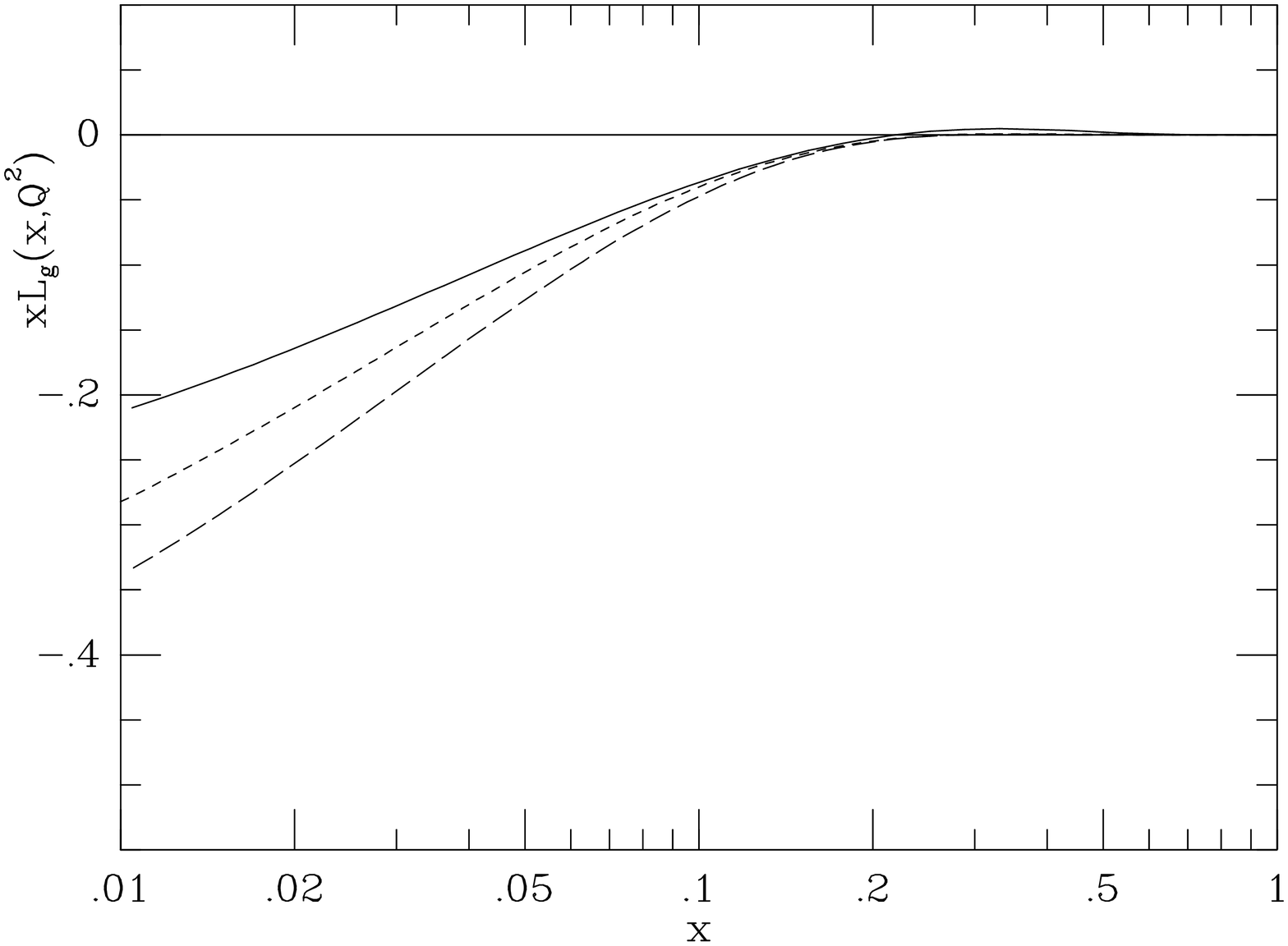,width=0.55\textwidth,angle=-90}}} 
\vskip 1cm
\centerline{\bf Figure 2} 
\vfill
%\newpage
\
\vfill 
\centerline{\protect\hbox{\psfig{file=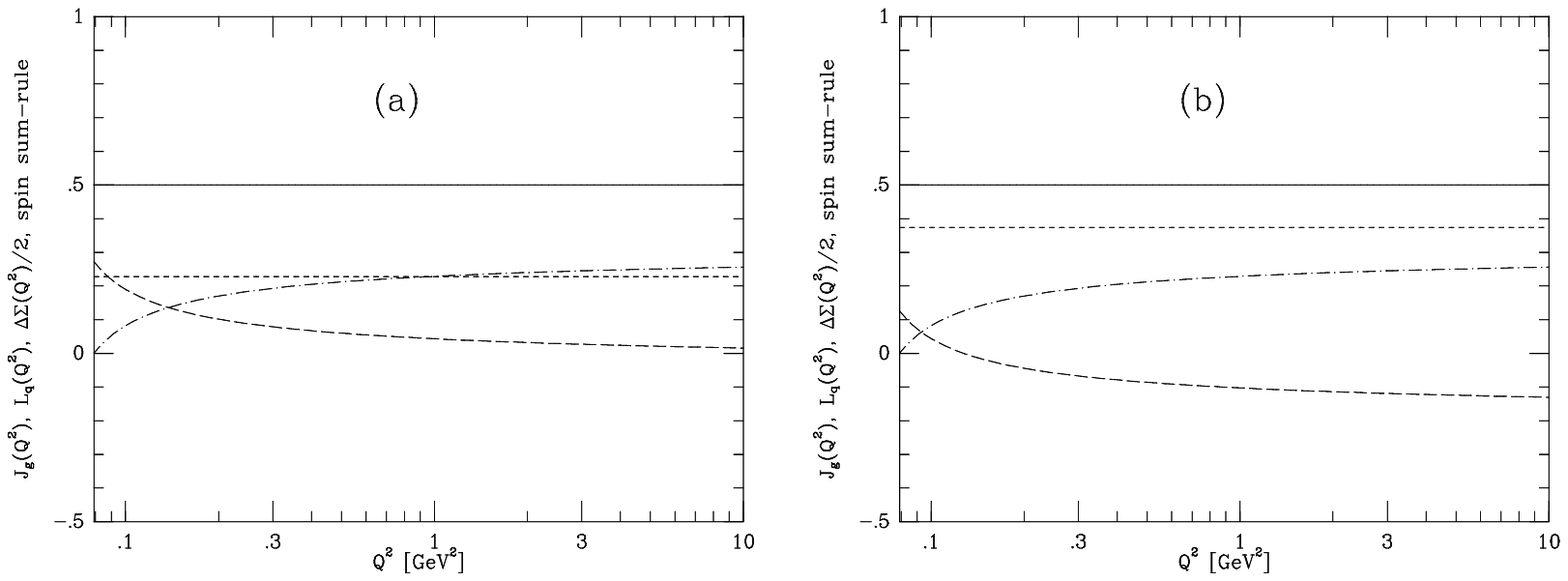,width=1.\textwidth}}} 
\vskip 1cm
\centerline{\bf Figure 3}
\vfill 

\begin{thebibliography}{99}

\bibitem{EMC88} 		EM Collaboration, J. Ashman et al., 
				Phys. Lett. B206 (1988) 364; 
				Nucl. Phys. B328 (1989) 1. 

\bibitem{SMC94} 		SM Collaboration, D. Adams et al., 
				Phys. Lett. B329 (1994) 399; 
                                erratum ibid. B339 (1994) 332; 
				Phys. Rev. D56 (1997) 5330.

\bibitem{ALTARELLI88} G. Altarelli, G.G. Ross, Phys. Lett. B212 (1988) 391.
	R.D. Carlitz, J.C. Collins and A.H. Mueller, Phys. Lett. B214 
	(1988) 229.
	
\bibitem{ALTARELLI97} G. Altarelli, R.D. Ball, S. Forte and G. Ridolfi, 
			Nucl. Phys. B496 (1997) 337. 

\bibitem{SEHGAL74} 	L.M. Sehgal, Phys. Rev. D10 (1974) 1663;

\bibitem{JAFFE90} 	R.L. Jaffe and A. Manohar, 
				Nucl. Phys. B337 (1990) 509.

\bibitem{LAMPE98} 	For a recent review: B. Lampe and E. Reya, 
				'Spin Physics and Polarized Structure 
                                 Functions', 
				hep-ph/9810270.

\bibitem{RATCLIFFE87}  P.G. Ratcliffe, Phys. Lett. B192 (1987) 180.
			
\bibitem{JI97} 		X. Ji, Phys. Rev. Lett. 78 (1997) 610;
				P. Hoodbhoy, X. Ji and W. Lu, Phys. Rev. D59,
                                (1999), 014013.

\bibitem{ja99}		S.V. Bashinsky and R.L. Jaffe, Nucl. Phys. B 536,  
		        (1998), 303.
\bibitem{JI96} 		X. Ji, J. Tang and P. Hoodbhoy, 
				Phys. Rev. Lett. 76 (1996) 740;
				P. H\"agler and A. Sch\"afer, 
				Phys. Lett. B430 (1998) 179;
				A. Harindranath and R. Kundu, 
                                Phys. Rev. D59, 116013, (1999).
                                O.E. Teryaev, hep-ph/9803403.

\bibitem{ji99}          P. Hoodbhoy, X. Ji and W. Lu, Phys. Rev. D 59,
                        074010, (1999).

\bibitem{SCOPETTA99} 	S. Scopetta and V. Vento, Phys. Lett. B 460 (1999) 8;
 			Erratum (to appear).

\bibitem{MA98} 		B.-Q. Ma and I. Schmidt, Phys. Rev. D58, 096008
                        (1999).

\bibitem{MA93} 		B.-Q. Ma and Q.-R. Zang, 
				Z. Phys. C58 (1993) 479;
				S.J. Brodsky and F. Schlumpf, 
				Phys. Lett. B329 (1994) 111;
				B.-Q. Ma, I. Schmidt and J. Soffer, 
                                Phys. Lett. B 441, (1998), 461.
				M.Wakamatsu and T. Watabe, hep-ph/9912500.

\bibitem{FACCIOLI99} 	M. Traini, P. Faccioli and V. Vento, 
				Few-Body Sys. Suppl. 11 (1999) 347;
				P. Faccioli, M. Traini and V. Vento,  
				Nucl. Phys. A656 (1999) 400.
\bibitem{CANO99} 		F. Cano, P. Faccioli and M. Traini, 
                                hep-ph/9902345.
\bibitem{METHOD1} 	M. Traini, L. Conci and U. Moschella,
				Nucl. Phys. A544 (1992) 731;
				M. Traini, V. Vento, A. Mair and A. Zambarda,
				Nucl. Phys. A614 (1997) 472 
                                and references therein.
\bibitem{METHOD2} 	A. Mair and M. Traini, Nucl. Phys. A624 (1997) 564;
				Nucl. Phys. A628 (1998) 296.
\bibitem{METHOD3} 	S. Scopetta, V. Vento and M. Traini, 
				Phys. Lett. B421 (1998) 64;
				Phys. Lett. B442 (1998) 28.

\bibitem{ISGUR78} 	N. Isgur and G. Karl, 
				Phys. Rev. D18 (1978) 4187; 
				D19 (1979) 2653; 
				D23 (1981) 817 (Erratum).

\bibitem{KEISTER91} 	B.D. Keister and W.N. Polyzou, 
				Adv. in Nucl. Phys. 20 (1991) 225;  
				F. Coester, 
				Prog. Part. Nucl. Phys. 29 (1992) 1;  
				J. Carbonell, B. Desplanques, 
                                V.A. Karmanov and J.-F. Mathiot, 
				Phys. Rep. 300 (1998) 215.
				S.J. Brodsky, H.-C. Pauli and S.S. Pinsky, 
				Phys. Rep. 301 (1998) 299.

\bibitem{PAOLO99} 	Paolo Longinotti,  Tesi di Laurea 1999, unpublished;
				M. Traini and P. Longinotti, in preparation.
\bibitem{TBM95}		M. Ferraris, M.M. Giannini, M. Pizzo, 
                        E. Santopinto and L. Tiator,
				Phys. Lett. B364 (1995) 231.
\bibitem{ROPELE95} 	M. Ropele, M. Traini and V. Vento, 
				Nucl. Phys. A584 (1995) 634.

\bibitem{MARTIN99} 	O. Martin, P. H\"agler and A. Sch\"afer,
                        Phys. Lett. B 448 (1999) 99.

\bibitem{BALITSKY97} 	I.I. Balitsky and X. Ji, 
				Phys. Rev. Lett. 79 (1997) 1225. 

\bibitem{MATHUR99} N. Mathur, S.J. Dong, K.F. Liu, L. Mankiewicz and 
			N. Mukhopadhyay, hep-ph/9912289.

\bibitem{BARONE98} V. Barone, T. Calarco and A. Drago, Phys. Lett. B431 (1998) 405.

\bibitem{HERMES99} HERMES Collaboration, A. Airapetian {\it et al},
hep-ex/9907020 (1999). 

\bibitem{BEYER98} 	M. Beyer, C. Kuhrts and H.J. Weber, 
                        Annals Phys. 269 (1998) 129. 
				
\end{thebibliography}
\end{document}